\documentclass[prl%
,superscriptaddress%
,twocolumn%
,amssymb%
]{revtex4}
\usepackage{dcolumn,epsfig}%

\newcommand{\bbbr}{{\mathbb R}}
\newcommand{\eps}{\varepsilon}
\newcommand{\length}{{\rm length}}
\newcommand{\perc}{{\rm perc}}
\newcommand{\sfrac}[2]{{\textstyle\frac{#1}{#2}}}

\begin{document}
\title{Scaling and Universality in Continuous Length Combinatorial
Optimization}
\date{\today}
\author{David Aldous}
\email{aldous@stat.berkeley.edu}
\affiliation{Statistics Department, University of California, Berkeley, California 94720--3860, USA} 
\author{Allon G.\ Percus} 
\email{percus@lanl.gov}
\affiliation{Computer \& Computational Sciences Division, Los Alamos
National Laboratory, Los Alamos, NM 87545--1663, USA}

\maketitle
{\bf 
We consider combinatorial optimization problems defined over random
ensembles, and study how solution cost increases when the optimal solution
undergoes a small perturbation $\delta$.  For the minimum spanning
tree, the increase in cost scales as $\delta^2$.  For the mean-field
and Euclidean minimum matching and traveling salesman problems in
dimension $d\ge 2$,
the increase scales as $\delta^3$; this is observed in Monte Carlo simulations in $d = 2,3,4$ and in theoretical analysis of a mean-field model.  We speculate that the scaling exponent
could serve to classify combinatorial optimization problems of this general kind into a small
number of distinct categories, similar to universality classes 
in statistical physics.
}

\vspace{0.18in}
\noindent
The interface of statistical physics, algorithmic theory, and
mathematical probability is an active research field, containing
diverse topics such as 
mixing times of Glauber-type dynamics
(\cite{DGM02} and many others),
reconstruction of broadcast information~\cite{EKPS00},
and probabilistic analysis of paradigm computational
problems such as $k$-SAT~\cite{CGHS03,MPZ02,MZKST}.
In this paper we introduce a new topic whose motivation is
simpler than those.

Freshman calculus tells us that, for a smooth function
$F: \bbbr\to\bbbr$ attaining its minimum at $x^*$, for $x$ near $x^*$
the relation between
$\delta = |x - x^*|$ and
$\eps = F(x) - F(x^*)$ is
$\eps \approx \frac{1}{2}F^{\prime \prime}(x^*) \delta^2$.
If instead we consider a function $F:\bbbr^d \to\bbbr$ on $d$-dimensional
space, sophomore calculus tells us that similarly
\[ \inf \{ F(x) - F(x^*) \ : \ |x - x^*| = \delta\}
\approx c \delta^2 \]
for appropriate $c$.
So in a sense the scaling exponent $2$ is
naturally associated with ``smooth'' or ``regular'' optimization
problems.

Now consider a graph-based combinatorial optimization problem, such as
the traveling salesman problem (TSP): each feasible
solution has $n$ constituents (edges)
and associated continuous costs (lengths), the sum of which gives
the overall solution cost.
Compare an arbitrary feasible solution ${\bf x}$ with the optimal (minimal)
solution ${\bf x}^*$ --- unique, for generic lengths --- by the two quantities
\begin{eqnarray*}
\delta_n({\bf x}) &=& \mbox{\{number of edges in {\bf x}
but not in {\bf x}$^*$\}}/n \\
\eps_n({\bf x}) &=& \mbox{\{cost difference between {\bf x}
and {\bf x}$^*$\}}/s(n)
\end{eqnarray*}
% $\delta_n({\bf x}) = \sfrac{1}{n} \times $
% (number of edges in ${\bf x}$ but not in ${\bf x}^*$)
% 
% $\eps_n({\bf x}) = \sfrac{1}{s(n)} \times $ (cost difference between ${\bf x}$
% and ${\bf x}^*$)
where $s(n)$ expresses the rate at which the optimal cost scales in $n$.
%In spin glass language, $\delta_n({\bf x})$ is one minus the overlap
%fraction between ${\bf x}$ and the ground state, and $\eps_n({\bf x})$
%is the rescaled energy difference between them.
Define $\eps_n(\delta)$ to be the minimum value of 
$\eps_n({\bf x})$ over all feasible solutions ${\bf x}$
for which $\delta_n({\bf x}) \geq \delta$.
Although the function $\eps_n(\delta)$ will depend on $n$ and
the problem instance, we anticipate that for typical instances
drawn from a suitable probability model % there will be in the
it will converge in the
$n \to \infty$ limit to some deterministic function
$\eps(\delta)$.

The {\em universality\/}
paradigm from statistical physics suggests there may be a scaling exponent $a$
defined by
\[ \eps(\delta) \sim \delta^a \mbox{ as } \delta \to 0 \]
and that the exponent should be robust under model details.
In statistical physics, universality classes typically refer to critical exponents that
characterize the behavior of measurable quantities both near and at a phase
a phase transition.  While $a$ is not a critical exponent here --- there
is no phase transition ---
we suggest that it could play a similar
role, categorizing combinatorial optimization problems
into a small set of classes.  If our analogy with freshman
calculus is apposite, we expect that the simplest problems should have
scaling exponent $2$.

This approach may seem obvious in retrospect, and fits within a
long-standing tradition in the physical sciences (see ``Discussion''
later).  However, it has never been proposed or explored explicitly.  In
this paper we report on three aspects of our program.  For the minimum
spanning tree (MST), a classic ``algorithmically easy'' problem solvable
to optimality by greedy methods,
we confirm that the scaling exponent is indeed $2$.
We then turn to two harder problems: minimum matching (MM) and the TSP.
Under a mean-field model, our new mathematical analysis methods combined
with numerics show that the scaling exponent is $3$ for both MM and TSP,
independent of the pseudo-dimension defined below.  For the Euclidean
model the exponent is $2$ in the (essentially trivial) one-dimensional
case, while Monte Carlo simulations suggest it is $3$ in higher
dimensions.

\vspace{0.14in}
\noindent
{\bf Models\/}

\vspace{0.09in}
\noindent
In the {\em Euclidean\/} model we take $n$ random points in a
$d$-dimensional cube whose volume scales as $n$.  Interpoint lengths are
Euclidean distances.  To reduce finite-size effects, we take
the space to have periodic (toroidal) boundary conditions when
calculating the distances.

In the {\em mean-field\/} or {\em random link\/} model we imagine 
$n$ random points in some abstract space such that the
${n \choose 2}$ vertex pair lengths are i.i.d.\ random variables
distributed as $n^{1/d} l$, with probability density $p(l) \sim l^{d-1}$
for small $l$.  Here $0 < d < \infty$ is the {\em pseudo-dimension\/}
parameter and the distribution of small single interpoint lengths  
mimics that in the Euclidean model of corresponding dimension $d$,
up to a proportionality constant.
Both models are set up so that nearest-neighbor distances are of order
$1$ and the scaling of overall cost in the optimization problems
is $s(n)=n$.

\vspace{0.14in}
\noindent
{\bf A simple case: the MST\/}

\vspace{0.09in}
\noindent
For the MST, for any reasonable model of interpoint lengths --- including the
two models above --- we expect a scaling exponent of $2$.  We will give
a rigorous account elsewhere~\cite{AS-AEU},
but the underlying idea is simple.  The classical greedy algorithm 
gives the following explicit {\em inclusion criterion\/} for whether
an edge $e = (v_1,v_2)$ of a graph belongs in the MST.
Consider the subgraph containing edges between any two vertices within
length $t$ of each other.
Let $\perc(e) \leq \length(e)$ be the smallest $t$ that keeps
$v_1$ and $v_2$ within the same connected component.
It is not difficult to see that
$e \in MST$ if and only if $\length(e) = \perc(e)$.

Given a probability model for $n$ random points and their interpoint
lengths, define a measure $\mu_n(x)$ on $x\in (0,\infty)$ in
terms of the expectation
\[
\mu_n(x) = \frac{1}{n} E \left| \{\mbox{ edges }e :
0 < \length(e) - \perc(e) < x\,\} \right|.
\]
For any reasonable model we expect an $n \to \infty$
limit measure $\mu(x)$, with a density $\nu(x)=d\mu/dx$ having
a non-zero limit $\nu(0^+)$.

% Given a probability model for $n$ random points and their interpoint
% lengths, now define a measure $\mu_n(A)$ on $A\subseteq [0,\infty)$ in
% terms of the expectation
% \[
% \mu_n(A) = \frac{1}{n} E \Biggl[\ \sum_{\mbox{vertex pairs }e}
% 1(\length(e) - \perc(e) \in A)\Biggr],
% \]
% where $1(\cdots)$ denotes the indicator function.
% This is equivalent to $1/2 \times$ the mean number of pairs per vertex that
% satisfy $\length(e) - \perc(e)\in A$.  We expect an $n \to \infty$
% limit measure $\mu(A)$, and since a spanning tree has average degree
% $\to 2$ as $n \to \infty$, the inclusion criterion implies that
% $\mu(0)=1$.  On $A\subseteq(0,\infty)$ it is
% reasonable to expect that $\mu(A)$ has a density function $\nu(x)$,
% $\mu(A)=\int_A\nu(x)\,dx$, with a nonzero limit $\nu(0^+)$.

% Given a probability model for $n$ random points and their interpoint
% lengths, now define a measure $\nu_n(\cdot)$ on $[0,\infty)$ in
% terms of the expectation
% \[
% \nu_n(\cdot) = \frac{1}{n} E \Biggl[\ \sum_{\mbox{vertex pairs }e}
% 1[\length(e) - \perc(e) \in \cdot]\Biggr],
% \]
% where $1[\cdots]$ denotes the indicator function.
% This is equivalent to $1/2$ the mean number of pairs per vertex that
% satisfy $\length(e) - \perc(e)\in\cdot$.  We expect an $n \to \infty$
% limit measure $\nu(\cdot)$, and since a spanning tree has average degree
% $\to 2$ as $n \to \infty$, the inclusion criterion implies
% that $\nu$ has mass $1$ at the point $0$.  On $(0,\infty)$ it is
% reasonable to expect that $\nu(\cdot)$ has a density function with a
% nonzero limit $\nu(0^+)\to c$.
% 
Now modify the MST by adding an edge $e$ with $\length(e) - \perc(e)
= b$, for some small $b$,
to create a cycle; then delete the longest edge $e^\prime \neq e$
of that cycle, which necessarily has $\length(e^\prime) = \perc(e)$.
This gives a spanning tree containing exactly one edge not in the MST
and having length greater by $b$.
Repeat this procedure with {\em every\/} edge $e$
for which $0<\length(e) - \perc(e) < \beta$, for some small $\beta$.
The number of such edges is $n\mu(\beta)\approx n\,\nu(0^+)\beta$
to first order in $\beta$, and as there is
negligible overlap between cycles, each of the new edges will
increase the tree length by $\sim\beta/2$ on average.  So
\[ \delta(\beta) \sim \nu(0^+)\beta, \quad 
\eps(\beta) \sim \nu(0^+)\beta^2/2 . \]
This construction must yield essentially the
minimum value of $\eps$ for given $\delta$, so
the scaling exponent is $2$.

\vspace{0.14in}
\noindent
{\bf Poisson Weighted Infinite Tree\/}

\vspace{0.09in}
\noindent
We now consider the
minimum matching (MM) and traveling salesman problem (TSP).
In MM, we ask for the minimum total length $L_n$ of
$n/2$ edges matching $n$ random points, and study the normalized limit
expectation
$\lim_{n \to \infty} \sfrac{2}{n} E[L_n]$.
Taking the mean-field model with $d = 1$ for simplicity,
the limit value $\pi^2/6$ was obtained in  ~\cite{MP85}
using the replica method from statistical physics.
We work in the framework of  ~\cite{me94}, which rederives this limit
rigorously by doing calculations within an $n = \infty$ limit structure,
the Poisson weighted infinite tree (PWIT).

Briefly, the PWIT is an infinite degree rooted tree
in which the edge weights (lengths) at each vertex are distributed as
the successive points $0 < \xi_1 < \xi_2 < \ldots$
of a Poisson process with a mean number $x^d$ of points in $[0,x]$,
i.e., a process with rate increasing as $d\,x^{d-1}$.
In this way, the PWIT corresponds to the mean-field model at
a given $d$ (see~\cite{me101} for review).

\begin{figure}
\epsfig{file=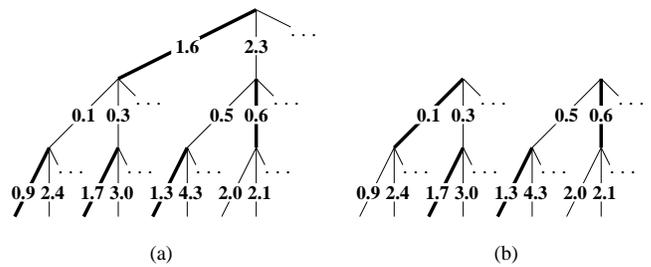,width=3.4in}
\caption{Matching on a PWIT, (a) with and (b)
without root node.  Numbers represent edge weights (lengths).}
\label{pwit}
\end{figure}

Consider a matching on an instance of a rooted PWIT, as well
as a matching on the same instance but with the root removed,
as shown in Fig.~\ref{pwit}.
Introduce the variable
\begin{eqnarray*}
X &=& 
\mbox{length of optimal matching on tree with root}\\
&-&
\mbox{length of optimal matching on tree without root}.
\end{eqnarray*}
Both lengths are infinite, so this is interpreted as a limit
of finite differences.
% reminiscent of uses of renormalization in statistical physics.
If $X_i$ is the analogous quantity for the $i$th constituent subtree of
the rootless PWIT instance and $\xi_i$ the length of the root's $i$th edge,
these variables satisfy the recursion
\begin{equation}
X = \min_{1 \leq i < \infty} (\xi_i - X_i).
\label{rec-1}
\end{equation}

Now take the $\{\xi_i\}$ to be the % Poisson($1$) on $(0,\infty)$,
Poisson-distributed edge lengths and the $\{X_i\}$ to be
independent random variables from the same random process that produces $X$.
Eq.~(\ref{rec-1}) is then a distributional equation for $X$, and can
be shown~\cite{me94} for $d = 1$ to have
as its unique solution the {\em logistic\/}
distribution
\begin{equation}
P(X \leq x) = \frac{1}{1+e^{-x}}, \quad -\infty < x < \infty .
\label{logistic}
\end{equation}

The PWIT structure further leads to the following {\em inclusion criterion\/}.
Consider an edge of length $x$ in the tree, and the two subtrees formed
by deleting
that edge.  The memoryless nature of the Poisson process allows us to
consider each of these subtrees as independent copies of a PWIT, with
their roots at the vertices of the deleted edge.  It may be seen that
including the edge in the optimal matching incurs a cost of $x-X_1-X_2$,
where $X_1$ and $X_2$ are the $X$ variables as defined above, but for
the two subtrees.  Thus, an edge of length $x$ is present in the minimal
matching if and only if
\begin{equation}
x < X_1 + X_2.
\label{inclusion}
\end{equation}

The probability density function for edge lengths in the minimal
matching is then
\[ f(x) = P(x<X_1+X_2), \quad 0 < x < \infty. \]
Here $X_1$ and $X_2$ are independent random variables distributed
according to Eq.~(\ref{logistic}), from which the mean edge length can be
calculated:
\[ \int_0^\infty x\, P(x<X_1+X_2)\,dx = \pi^2/6. \]

\vspace{0.14in}
\noindent
{\bf Mean-field MM and TSP\/}

\vspace{0.09in}
\noindent
The previous section summarized analysis from  ~\cite{me94};
now we continue with new analysis.
To study scaling exponents, we introduce a parameter $\lambda > 0$
that plays the role of a Lagrange multiplier.
Penalize edges used in the optimal matching by adding $\lambda$ to their
length.  Let us study optimal solutions to the MM problem on this new
penalized instance.  Precisely, on a realization of the PWIT, 
define $Y$ and $Z$ as
\begin{eqnarray*}
  && 
\mbox{length of optimal matching on new tree with root}\\
&-&
\mbox{length of optimal matching on new tree without root}
\end{eqnarray*}
where $Y$ and $Z$ differ in the definition of the edge lengths of the new tree:
for $Y$, the edges penalized are those employed by the original {\em
rooted\/} optimal matching; for $Z$, they are those employed by the
original {\em rootless\/} optimal matching.

For the penalized problem the recursion Eq.~(\ref{rec-1}) for $X$
is supplemented by the following recursions for $(X,Y,Z)$
jointly.  Let $i^*$ be the value of $i$ that minimizes $\xi_i - X_i$.
Then
\begin{eqnarray*}
Y&=& \min_i (\xi_i - (Z_i+\lambda)\,1(i=i^*) - Y_i\,1(i \neq i^*)) \\
Z&=& \min_i (\xi_i - Y_i)
\end{eqnarray*}
where, as before, the $\{Y_i\}$ and $\{Z_i\}$ are independent random
variables from the same random process producing $Y$ and $Z$.

Moreover, we get an inclusion criterion, analogous to Eq.~(\ref{inclusion}):
an edge of length $x$ is included if and only if
\begin{eqnarray*}
x + \lambda < Z_1 + Z_2 &&\ \mbox{ if edge used in optimal matching}\\
x < Y_1 + Y_2 &&\ \mbox{ if edge not used in optimal matching}. 
\end{eqnarray*}
In terms of the expected unique joint distribution for
$(X,Y,Z)$,
the quantities $\delta$ and $\eps$ that compare the
penalized solution (as a {\em non-optimal\/} solution of the
original problem) with the original optimal solution are
\begin{eqnarray*}
\delta(\lambda) &=&
\int_0^\infty 
P\{\mbox{edge of length $x$ is in optimal penalized}\\
&&\qquad\mbox{matching but not in optimal matching}\}\,dx\\
&=& \int_0^\infty P(X_1+X_2 < x < Y_1+Y_2)\,dx
\end{eqnarray*}
and
\begin{eqnarray*}
\eps(\lambda) &=&
\int_0^\infty 
x\,P\{\mbox{edge of length $x$ is in optimal penalized}\\
&&\qquad\mbox{matching}\}\,dx - \pi^2/6 \\
&=& \int_0^\infty
x\,[P(x < X_1+X_2,\ x < Z_1+Z_2-\lambda) \\
&&\qquad +\ P(X_1+X_2 < x < Y_1+Y_2)]\,dx - \pi^2/6 .
\end{eqnarray*}
By the theory of Lagrange multipliers these functions $\eps(\lambda),
\delta(\lambda)$ determine $\eps(\delta)$.
We do not have explicit analytic expressions analogous to Eq.~(\ref{rec-1}) for the joint
distribution of $(X,Y,Z)$ in terms of $\lambda$.  However,
we can use routine bootstrap Monte Carlo simulations
to simulate the distribution and thence estimate the functions
$\delta(\lambda)$ and $\eps(\lambda)$ numerically.
And as indicated in  ~\cite{me94} sec.\ 6.2 and \cite{KM87,Cerf},
the mean-field MM and the mean-field TSP can be studied using similar
techniques; the TSP analysis is just a minor variation of the
MM analysis.
For instance, recursion Eq.~(\ref{rec-1}) becomes
\[ X = {\min_{1 \leq i < \infty}}^{[2]} (\xi_i - X_i) \]
where $\min^{[2]}$ denotes second-minimum.

Table~\ref{bootstrap} reports numerical results showing
good agreement with $\eps \propto \delta^3$ in both problems
for $d=1$.  These numerics are compatible with independent MM results
obtained recently~\cite{Ratieville}, as well as with our direct
simulations on mean-field TSP instances at $n=512$.  The same exponent
$3$ arises for other $d$.

\begin{table}[b]
\caption{Scaling for mean-field MM and TSP in pseudo-dimension $d=1$,
obtained by simulating joint distribution of $(X,Y,Z)$.
Results show a good fit to $\eps \sim 2.3 \delta^3$
and $2.0 \delta^3$.  In more detail, $\delta$ scales as $\lambda^{1/2}$
while $\eps$ scales as $\lambda^{3/2}$.
Estimates for $\eps$ have standard deviation about $0.001$ 
for MM and $0.003$ for TSP.}
\vspace{0.1in}
\begin{tabular}{cc|ccccccc|cccccc}
\toprule
&&&&&MM&&&&&&&TSP&&\\
\hline
$\lambda$&&&$\delta$&&$\eps$&&$2.3\delta^3$&&&$\delta$&&$\eps$&&$2.0\delta^3$\\
\hline
0.02 &&& 0.112 && 0.004 && 0.003 &&& 0.128 && 0.009 && 0.006\\
0.04 &&& 0.156 && 0.010 && 0.009 &&& 0.175 && 0.015 && 0.011\\
0.06 &&& 0.190 && 0.017 && 0.016 &&& 0.212 && 0.023 && 0.019\\
0.08 &&& 0.219 && 0.024 && 0.024 &&& 0.243 && 0.030 && 0.029\\
0.10 &&& 0.244 && 0.035 && 0.033 &&& 0.270 && 0.042 && 0.039\\
0.12 &&& 0.267 && 0.042 && 0.044 &&& 0.300 && 0.053 && 0.051\\
0.14 &&& 0.287 && 0.053 && 0.054 &&& 0.318 && 0.065 && 0.064\\
0.16 &&& 0.306 && 0.067 && 0.066 &&& 0.340 && 0.077 && 0.079\\
0.18 &&& 0.323 && 0.080 && 0.078 &&& 0.360 && 0.091 && 0.093\\
0.20 &&& 0.340 && 0.089 && 0.090 &&& 0.379 && 0.104 && 0.109\\
\botrule
\end{tabular}
\label{bootstrap}
\end{table}

\vspace{0.14in}
\noindent
{\bf Euclidean MM and TSP\/}

\vspace{0.09in}
\noindent
We consider the $d=1$ case where the scaling exponent can
be found exactly, and give numerical results for other cases.  We
restrict the discussion to the Euclidean TSP, although
as for the mean-field model, MM is phenomenologically similar.

\begin{figure}
\epsfig{file=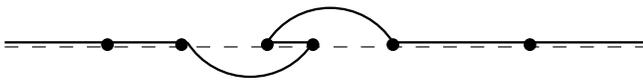,width=3.4in}
\caption{``2-change'' schematic.  Original optimal
tour is shown by dashed line.  New optimal tour on penalized instance is shown
by solid line: over sufficiently short lengths, tour doubles back to avoid
using penalized edges.}
\label{2change}
\end{figure}

Take the Euclidean TSP in $d=1$, with periodic boundary conditions.
The optimal tour here is trivial (with high probability a straight line
of length $n$) but nevertheless instructive to analyze.  As before,
add a penalty term $\lambda$ to each edge used in the tour,
and consider how the optimal tour changes in this new penalized instance.
When $\lambda$ is small, changes to the tour will consist of
``2-changes'' shown in Fig.~\ref{2change}, and will occur when
an original edge length is less than $\lambda$.  A simple nearest-neighbor
distance argument gives the distribution of edge lengths in the
original tour as $p(l)\sim e^{-l}$.  Since two edges are modified in
each 2-change,
\[ \delta(\lambda)=2\int_0^\lambda p(l)\,dl\sim 2\lambda,
\quad \eps(\lambda)=2\int_0^\lambda l\,p(l)\,dl\sim\lambda^2. \]
The scaling exponent of 2 is not surprising, as the 1-d TSP behaves
very similarly to the 1-d MST on penalized instances.  Furthermore,
it is consistent with the intuition that the
``easiest'' problems scale in this way.
A similar argument applies to MM, and in both cases a more rigorous
analysis yields bounds on $\delta$ and $\eps$ that confirm the exponent.
% A similar argument applies to the MM.  In both cases, a more rigorous
% analysis yields bounds on $\delta$ and $\eps$ that confirm the scaling
% exponent 2, consistent with the intuition that the ``easiest'' problems
% scale in this way.
% A similar argument applies to the MM, and in both cases a more rigorous
% analysis yields bounds on $\delta$ and $\eps$ that confirm the scaling
% exponent 2.  The value is consistent with the intuition that the
% ``easiest'' problems scale in this way.

\begin{figure}
\epsfig{file=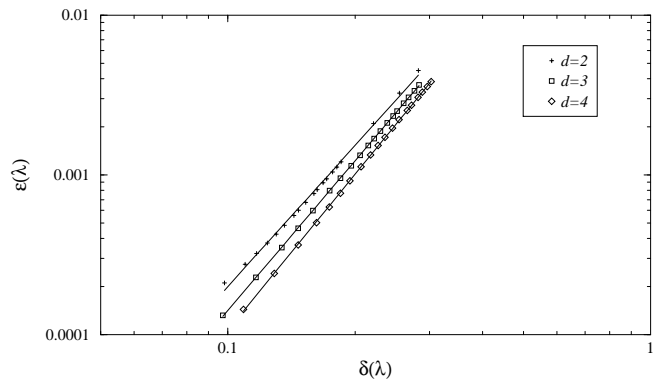,width=3.4in}
\caption{Scaling for Euclidean TSP in $d=2,3,4$, based on
exact solutions for 100 instances in each case.
Data points correspond to $\lambda$ values from 0.004 to 0.05.
Slopes of best-fit lines vary from 2.94 to 3.24.
Standard deviation is about $3\times 10^{-3}$ for $\delta$ and
$3 \times 10^{-5}$ for $\eps$.}
\label{simulations}
\end{figure}

For $d>1$, numerical results are shown in Fig.~\ref{simulations}.  These
have been obtained by finding exact solutions to randomly generated
$n=512$ Euclidean instances in $d=2,3,4$,
using the {\em Concorde\/} TSP solver~\cite{concorde}.  For each
instance, the optimum was obtained on the original instance as well as
on the instance penalized with a range of $\lambda$ values.  For
each $\lambda$ value, $\delta(\lambda)$ and $\eps(\lambda)$ were
averaged over the sample of instances.
The resulting numerics are closely consistent
with a scaling exponent of 3 (in spite of suffering from some finite-size
effects at smaller $\delta$), suggesting that the mean-field picture gives
the correct exponent in all but the trivial 1-dimensional case.  In the
language of critical exponents, this would correspond to an
``upper critical dimension'' of 2.

\vspace{0.14in}
\noindent
{\bf Discussion\/}

\vspace{0.09in}
\noindent
The goal of our scaling study has been to address a new kind of
problem in the theory of algorithms, using concepts from statistical
physics.
% use concepts from
% statistical physics in order to suggest and address new problems in the
% theory of algorithms.
% Our scaling results complement a rich array of existing results
% by computer scientists, mathematicians and physicists alike.  
Traditionally, work on the TSP within the theory of algorithms~\cite{TSPbook}
has emphasized algorithmic performance, rather than the kinds of
questions we ask here.  Rigorous study of the Euclidean TSP model within
mathematical probability~\cite{steele97} has
yielded a surprising amount of qualitative information: existence
of an $n \to \infty$ limit constant giving the mean edge-length in 
the optimal TSP tour~\cite{BHH}, and large deviation bounds for the
probability that the total tour length differs substantially from its
mean~\cite{RT89}.  However, calculation of explicit constants in
dimensions $d \geq 2$ seems beyond the reach of analytic techniques.
For the mean-field bipartite MM problem, impressive recent
work~\cite{LW03,NPS03} has proven an exact formula giving the
expectation of the finite-$n$ minimum total matching length, though such
exact methods seem unlikely to be widely feasible.

On the other hand, there has been significant progress over the past
twenty years in the use of statistical physics techniques on
combinatorial optimization problems in general.  Finding optimal
solutions to these problems is a direct analog to determining ground
states in statistical physics models of disordered
systems~\cite{Moore87}.  This observation has motivated the development
of such approaches as simulated annealing~\cite{KGV83}, the replica
method~\cite{MPV87} and the cavity method~\cite{MPZ02}.  Condensed
matter physics, and particularly models arising in spin glass theory,
has provided a powerful means to study algorithmic problems; at
the same time, algorithmic results have implications for the
associated physical models.  It is instructive to consider our work in
that context.

Researchers in the physical sciences have long been interested in the
low-temperature thermodynamics~\cite{MPV87,Dotsenko} of disordered
systems, investigating properties of near-optimal states in spin glass
models.  Our procedure for studying
near-optimal solutions by way of a penalty parameter is similar to a
method, known as $\epsilon$-coupling~\cite{eps,MPR-T}, used for
calculating low-energy excitations in spin glasses.  Making use of
this method, physicists have obtained quantities
closely analogous to our scaling exponents for models of RNA
folding~\cite{MPR-T}.  Furthermore, in the last year independent
work~\cite{Ratieville} has explored $\epsilon$-coupling on MM,
numerically identifying a different but related scaling exponent.

For the TSP, analytical and numerical studies have been performed over
fifteen years ago~\cite{MP86,Sourlas} on the thermodynamics of the
model, with overlap quantities calculated for near-optimal solutions.
The results have suggested that at low temperature $T$, the cost excess
$\eps$ scales as $T^2$ while the average fraction of differing edges
between solutions ($1-q$, with $q$ being the ``overlap fraction'')
scales as $T$.  This leads to $\eps\sim (1-q)^2$, in apparent
contradiction with our exponent of 3.  However, at low temperatures, $q$
represents overlaps between {\em typical} near-optimal solutions,
whereas our $\delta$ measures overlaps between a near-optimal solution
and the optimum.  The different definitions of these two quantities
could account for the discrepancy in scaling exponent: it is not
surprising that $1-q$ grows faster than $\delta$ as one considers
solutions of increasing cost.  At the same time, a possible implication
of these results is that at low temperature, $\delta\sim T^{2/3}$.  We
are not aware of any direct theoretical arguments to explain this, and
consider it an intriguing open question.

It is also important to note that the underlying property $\delta \to 0$
as $\eps \to 0$ cannot always be taken for granted.  This property is
called {\em asymptotic essential uniqueness\/} (AEU)~\cite{me94}.  AEU
requires, among other things, that the optimum itself be unique.  In
principal, even if it is not, one could still analyze near-optimal
scaling by considering sufficiently local perturbations from a given
optimum.  It is natural to expect the resulting exponent to be
independent of the specific optimum chosen.  However, this may {\em
not\/} be true in the event of what statistical physicists call {\em
replica symmetry breaking\/} (RSB)~\cite{MPV87,Dotsenko}.  AEU is a
special case of replica symmetry, so while RSB implies the absence of
AEU, the absence of AEU does not necessarily imply RSB.  A current
debate in the condensed matter literature concerns whether or not
low-temperature spin glasses display RSB~\cite{eps,KM}.  It is generally
believed that RSB is incompatible with unique non-zero values of various
scaling exponents.  Thus, the correct approach to analyzing near-optimal
scaling in such problems remains another open question.

One final example may serve to illustrate the diversity of possible
applications for our type of scaling analysis, as well as an instance
where the absence of AEU is surmountable.
In oriented percolation on the two-dimensional lattice, 
there are independent random traversal times on each oriented
(up or right) edge.  The percolation time $T_n$ is the minimum,
over all ${2n \choose n}$ paths from $(0,0)$ to $(n,n)$,
of the time to traverse the path.
So $(2n)^{-1} E[T_n] \to t^*$, a time constant.
It is elementary that there will be near-optimal paths,
with lengths $T^\prime_n$ such that $n^{-1}(ET^\prime_n - ET_n) \to 0$
and which are almost disjoint from the optimal path.
So our $\eps(\delta)$ analysis applied to paths will not 
be useful: even with a unique optimum, AEU will not hold.
But we can rephrase the problem in terms of flows.
A {\em flow\/} on the $n \times n$ oriented torus assigns
to each edge a flow of size $\in [0,1]$, such that at each
vertex, in-flow equals out-flow.
Let $t(\delta)$ be the minimum, over all flows with
mean flow-per-edge $= \delta$, of the flow-weighted average
edge traversal time.
In the $n \to \infty$ limit, one can show that as $\delta\to 0$,
$t(\delta)\to t^*$ where $t^*$ is the same limiting constant as
before.
We therefore expect a scaling exponent
$t(\delta) - t^* \sim \delta^a$.
Mean field analysis~\cite{meFPP} gives $a = 2$, and
Monte Carlo study of the $d = 2$ case is in progress.

\vspace{0.14in}
\noindent
{\bf Conclusions\/}

\vspace{0.09in}
\noindent
We have studied the scaling of the relative cost difference $\eps$ between
optimal and near-optimal solutions to combinatorial optimization problems,
as a function of the solution's relative distance $\delta$ from optimality.
This kind of scaling study, although well accepted in theoretical physics,
is new to combinatorial optimization problems.
For the MST, we have found $\eps\sim\delta^2$.  For the MM and TSP,
in the 1-d Euclidean case $\eps\sim\delta^2$ as well, while in both
the mean-field model and higher Euclidean dimensions
$\eps\sim\delta^3$.

The scaling exponent may categorize combinatorial optimization problems
into a small number of classes.  The fact that MST is solvable by a
simple greedy algorithm, and that the 1-d case of the MM and TSP is
essentially trivial, suggests that a scaling exponent of 2 characterizes
problems of very low complexity.  The exponent of 3 characterizes
problems that are algorithmically more difficult.  Of course, this is a
different kind of classification from traditional notions of
computational complexity: MM is solvable to optimality in $O(n^3)$ time
whereas the TSP is NP-hard.  Rather, these exponent classes are
reminiscent of universality classes in statistical physics, which unite diverse
physical systems exhibiting identical behavior near phase transitions.

A key question in the study of critical phenomena is whether mean-field
models correctly describe phase transition behavior in the geometric
models they approximate.  The TSP and MM do not involve critical
behavior, but the fact that mean-field and geometric scaling exponents
coincide for $d\ge 2$ is significant.  It provides evidence that in a
combinatorial setting, the mean-field approach can give a valuable and
accurate description of the structure of near-optimal solutions.

\noindent
{\footnotesize We thank Mike Steele for helpful discussions.
David Aldous's research is supported by NSF Grant DMS-0203062.
Allon Percus acknowledges support from DOE LDRD/ER 20030137, as well as
the kind hospitality of the UCLA Institute for Pure and Applied
Mathematics where much of this work was conducted.
}

\end{document}